\documentclass[a4paper]{jpconf}
\usepackage{graphicx}
\usepackage{amssymb, amsmath, stmaryrd, latexsym, textcomp, wasysym,  ifsym}
\begin{document}
\title{Effect of Crystal Growth Rate on Liquid-like Droplets Formation in the hcp Solid Helium}

\author{N P Mikhin, A P Birchenko, 
A S Neoneta, E Ya Rudavskii and Ye~O~Vekhov}

\address{Quantum Fluids and Solids Department, B.Verkin Institute for Low Temperature Physics and Engineering, National Academy of Sciences of Ukraine, 
\\ 47 Lenin Ave., Kharkov 61103, Ukraine}

\ead{yegor\_v@ukr.net}

\begin{abstract}
The samples of hcp solid helium (1\% $^3$He in $^4$He) are studied by NMR technique. The samples are grown by the blocking capillary method at different growth rates (about 8, 2, and 0.08~mK/s). The NMR technique is used for phase identification by measurements of the diffusion coefficient $D$ and the spin-spin relaxation time $T_2$ at temperatures of $1.3-2.0$~K and pressures of $34-36$~bar. Along with $D$ and $T_2$ for the hcp phase, we simultaneously observe the $D$ and $T_2$ typical for a liquid at growth rates 8 and 2~mK/s. It means that liquid-like inclusions are quenched from the melting curve during fast crystallization of the samples. It is also shown that the slower growth rate corresponds to a smaller size of liquid-like droplets. It  results from lower spatially restricted values of $D$ and, finally, absence of these inclusions at the longest crystallization times. The diffusion coefficient measured for liquid-like droplets is also decreasing during the NMR experiment at constant temperature which indicates the reduction of the size of these droplets. Liquid-like droplets are shown to disappear after sample annealing near the melting curve.
\end{abstract}

\section{Introduction}

The measured diffusion coefficient $D$ of solid 1\% $^3$He-$^4$He samples with coexisting bcc and hcp phases \cite{Mikhin.2001} and the pressure drop occurring when the bcc-hcp interphase in $^4$He disappears \cite{Mikhin.2007} suggest that non-annealed solid helium samples contain a large number of liquid-like inclusions at the bcc-hcp interphase. The recent visual observation of the defect structure in solid $^4$He \cite{Sasaki.2008} has shown that single-phase samples grown by the blocking capillary method are polycrystals with grains of micrometer size and a liquid phase can be presented at grain boundaries. The observation \cite{Sasaki.2008} were supported by NMR measurements of the coefficient $D$ \cite{Vekhov.2010} which show that the macroscopic liquid-like inclusions can exist in samples for quite a long time under high pressure even at the absence of equilibrium interphases (in a hcp sample). This study continues logically the previous investigations and supplements them with measuring the spin-spin relaxation time in a hcp sample of solid helium in addition to the diffusion coefficient. The effect of the growth rate of a crystal upon the size and properties of such liqiud-like inclusions have been investigated for the first time.

\section{Experimental technique}

\begin{figure}[h]
\includegraphics[width=22pc]{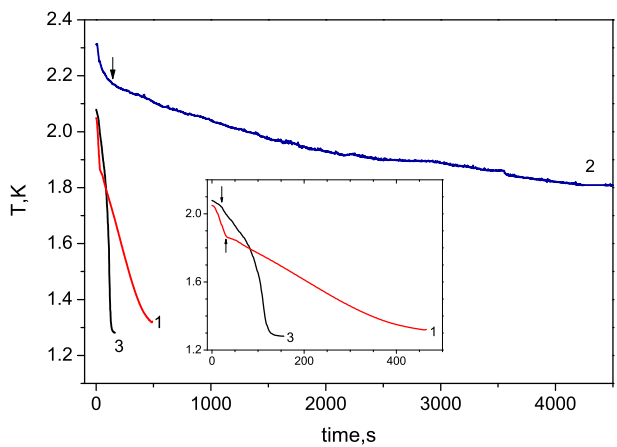}\hspace{2pc}%
\begin{minipage}[b]{14pc}\caption{\label{fig_growth-rate}(Color online) Kinetics of temperature variations in a course of helium crystals growth at different cooling rates. Arrows mark the onset of the crystallization. Curves 1 and 2 --- cooling rates of this study, curve 3 --- cooling rate of \cite{Vekhov.2010}. Insert: curves 1, 3 on a large scale.}
\end{minipage}
\end{figure}

Helium is fed to the cavity of an experimental NMR cell which is cylinder 16~mm long and 8~mm in diameter. A NMR coil is wound around the cavity (sample). The measurements are maid in a temperature interval of $1.3-2.0$~K. The used hcp crystals (1\% $^3$He-$^4$He) correspond to the pressures $\AC 35\pm 1$~bar ($V_m\AC 20.2\pm 0.1$~cm$^3$/mol), i.e. they are grown from a normal liquid above the upper triple point (bcc-hcp-He~I). The obtained crystals may be divided arbitrary into two types. The crystals of type 1 are prepared by fast crystallization at a rate of $\AC 2$~mK/s (see Fig.1, curve 1) which produced a grate quantity of defects in the sample. The type 2 crystals are prepared using a thermostabilization system which allows a slow growth of a crystal at a cooling rate of  $\AC 0.08$~mK/s (see Fig.1, curve 2). Note that in \cite{Vekhov.2010} the samples are grown at even a higher cooling rate of  $\AC 8$~mK/s (see Fig.~\ref{fig_growth-rate}, curve 3).

To identify the phase composition of samples obtained at different growth rates, NMR measurements were performed at a frequency of $f_0=9.15$~MHz using the standard Carr-Pursell \cite{Carr.1954} and Hahn (stimulated echo) \cite{Hahn.1950} techniques. The techniques used allowed measuring the spin-spin relaxation time $T_2$ and the spin diffusion coefficient $D$ of the $^3$He atoms in the $^4$He matrix.

\section{Single-phase samples: hcp and bulk liquid helium}

\begin{figure}[h]
\begin{center}
\begin{minipage}[t]{0.475\linewidth}
\includegraphics[width=1\linewidth]{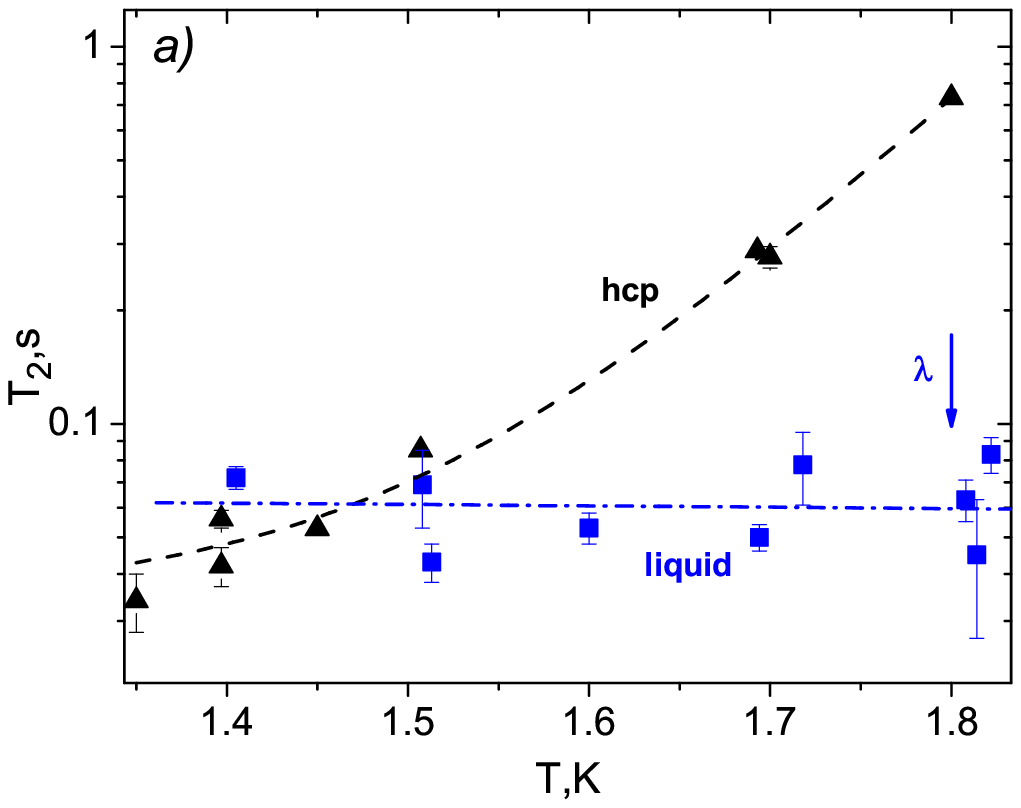}
\end{minipage}
\hfill 
\begin{minipage}[t]{0.505\linewidth}
\includegraphics[width=1\linewidth]{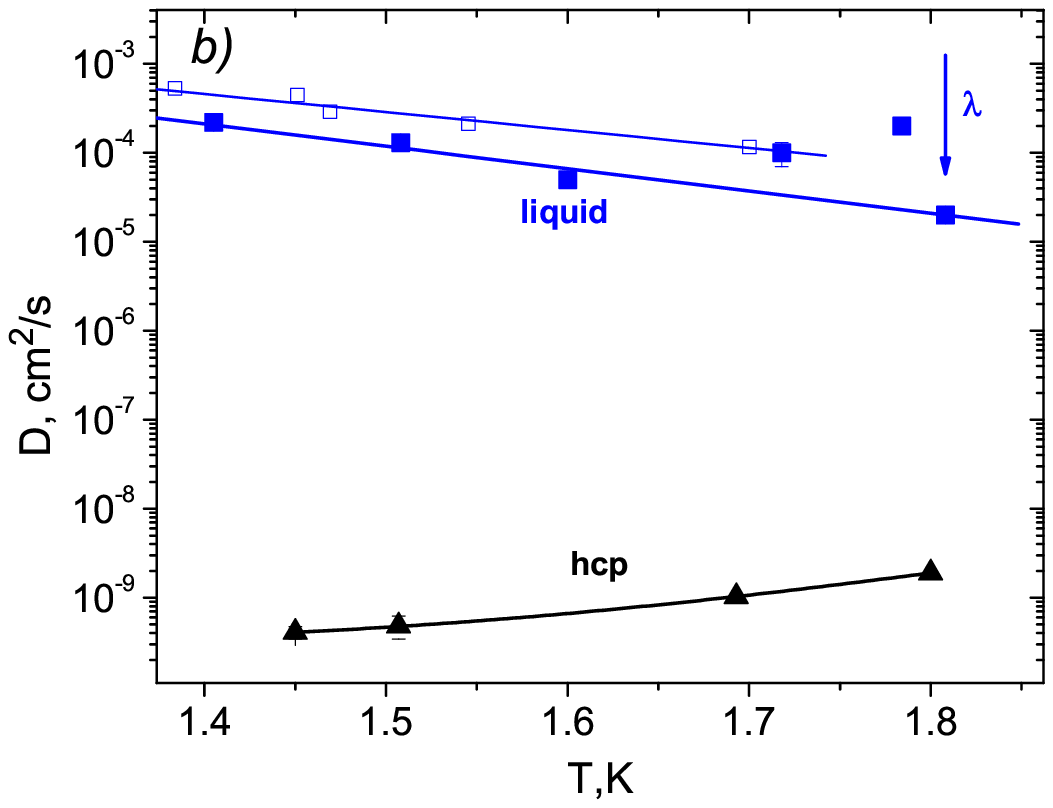}
\end{minipage}
\caption{(Color online) The temperature dependences of the spin-spin relaxation time  $T_2$ (a) and the diffusion coefficient $D$ (b) for an annealed single-phase hcp sample at $P\AC 35$~bar ($\blacktriangle$) and a bulk helium liquid at $P\AC 25$~bar (blue $\blacksquare$) and $P\AC 19$~bar (blue $\square$) \cite{Garwin.1959}. Arrows mark the temperature of the $\lambda$-transition in liquid helium. The solid, dashed, and dash-dot lines are over experimental points.}
\label{fig_T2_Donephase}
\end{center}
\end{figure}

At first calibration NMR experiments were carried out to investigate the temperature behavior of the time $T_2$ and the diffusion coefficient $D$ in single-phase samples: a slowly grown and well annealed hcp crystal at $P\AC 35$~bar and a bulk liquid at $P\AC 25$~bar, i.e. slightly below the melting curve. The measurement results are shown in Fig.~\ref{fig_T2_Donephase}. It is seen that the coefficient $D$ of the hcp phase and the liquid differ by several orders of magnitude, while the times $T_2$ vary only slightly, especially at $T<1.6$~K.

\section{Two-phase samples: liquid-like inclusions in hcp matrix}

\begin{figure}[h]
\begin{center}
\begin{minipage}[t]{0.475\linewidth}
\includegraphics[width=1\linewidth]{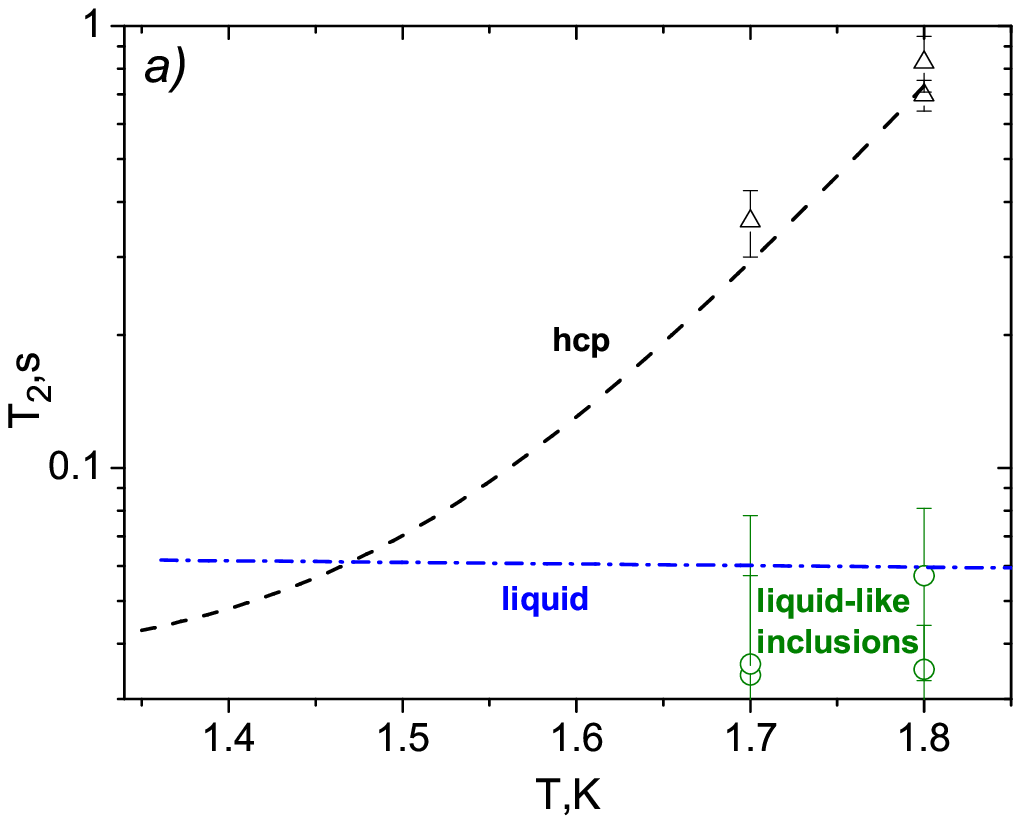}
\end{minipage}
\hfill 
\begin{minipage}[t]{0.505\linewidth}
\includegraphics[width=1\linewidth]{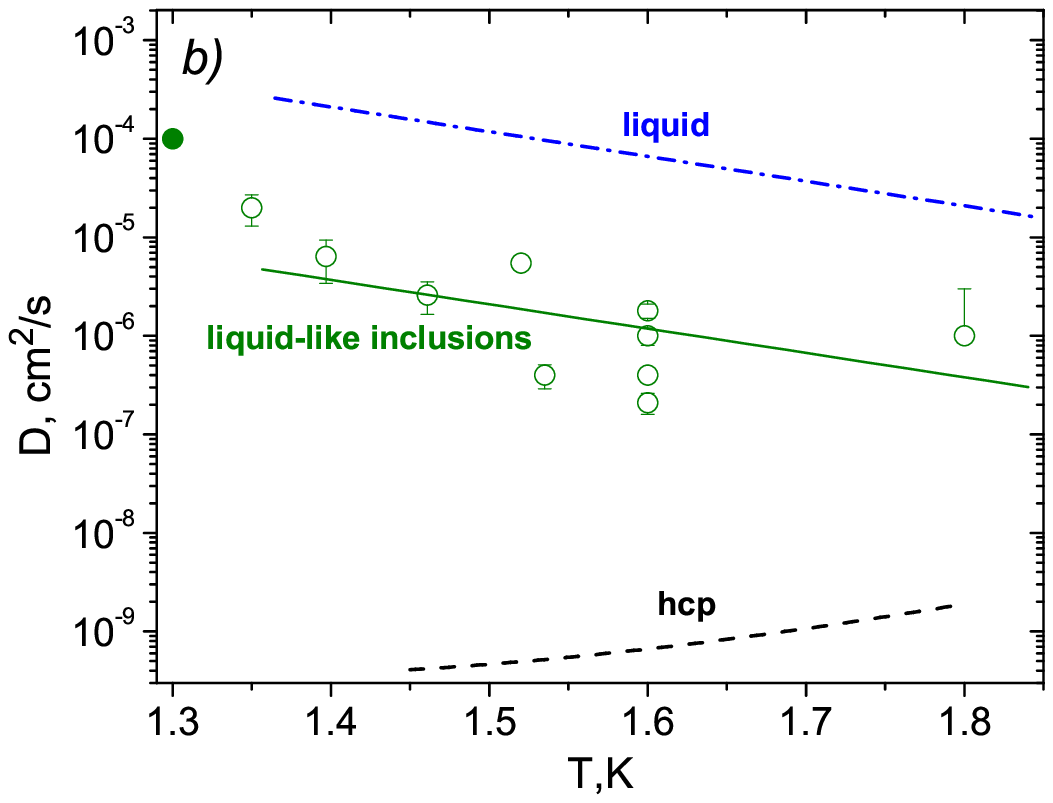}
\end{minipage}
\caption{(Color online) The temperature dependence of (a) the spin-spin relaxation time $T_2$ of (green $\Circle$) metastable liquid-like inclusions (growth rate $\AC 2$~mK/s), ($\triangle$) the hcp phase in the presence metastable inclusions and (b) the diffusion coefficient $D$ of (green $\Circle$)  metastable liquid-like inclusions (growth rate $\AC 2$~mK/s), (green $\CIRCLE$) metastable liquid-like inclusions (growth rate $\AC 8$~mK/s)  \cite{Vekhov.2010}. Dashed line --- single phase hcp sample at $P\AC 35$~bar, dash-dot line --- bulk liquid at $P\AC 25$~bar (see Fig.~\ref{fig_T2_Donephase}).}
\label{fig_T2_D_liqincls}
\end{center}
\end{figure}

The investigations of the time $T_2$ and the diffusion coefficient $D$ on solid helium samples grown at the cooling rate $\AC 2$~mK/s show that the NMR signal has two contributions --- one from the hcp phase and the additional contribution up to 30\%. The time $T_2$ of the additional contribution  was approximately equal to $T_2$ of single-phase samples consisting of only bulk liquid.\footnote{Note that an additional contribution to $T_2$ was observed only at $T=1.7-1.8$~K, i.e. in the region where the times of spin-spin relaxation of the hcp phase and bulk liquid helium differ considerably.} The coefficient $D$ of this contribution was about $1.5-2$ orders of magnitude lower than in the bulk liquid (see Fig.~\ref{fig_T2_D_liqincls}). The results in Fig.~\ref{fig_T2_D_liqincls} show that the liquid droplets were captured during crystallization on fast cooling. The lower diffusion coefficient in the liquid-like inclusions may be due to the restricted character of the coefficient $D$ (during the NMR measurement the $^3$He atom reaches the wall of the droplet, which reduces the value of the $D$, e.g., see Ref.~\cite{Wayne.1966}). 

Note that the additional contribution  to the NMR signal was unobservable, within the technique sensitivity ($\pm 3\%$), in samples grown at a rate $\AC 0.08$~mK/s.

Fig.~\ref{fig_T2_D_liqincls} also illustrates the experimental point (green $\CIRCLE$) obtained \cite{Vekhov.2010} for a sample grown at the cooling rate  $\AC 8$~mK/s. In this case the $D$-magnitude exceeds  the corresponding values for liquid-like inclusions formed at the cooling rate $\AC 2$~mK/s and approaches the $D$-coefficient of a bulk liquid. This may indicate that the inclusions formed at the rate $\AC 8$~mK/s are large in size than those in the samples obtained at the growth rate $\AC 2$~mK/s. In other words, the higher is the cooling rate during the sample growth, the larger liquid-like droplets can be obtained.

The NMR measurement on a fast-grown sample also showed a prolonged monotonic decrease in $D$ (almost an order of magnitude during four hours) of liquid-like inclusions  at constant $T=1.60$~K. This is presumably because the size of the inclusions reduces gradually on annealing. This kind of kinetics is another evidence in favor of the restricted character of diffusion in liquid-like inclusions.

It is appropriate to mention here occasional fast and drastic  changes that can occur spontaneously in the properties of liquid-like inclusions. The spin-spin relaxation time falls by an order of magnitude in comparison with $T_2$ of liquid-like inclusions. The diffusion coefficient becomes lower or nearly equal to $D$ of the hcp phase. These changes may be a manifestation of a transition to a disordered state, which calls, however, for further experimental investigation.

Finally, no additional contribution of liquid-like inclusions was observed after annealing (holding two-phase samples near the melting curve at $T\AC 1.95$~K for about one or two hours). This fact points to the metastable character of the revealed liquid-like inclusions.

In pursuing the goal of identifying the phase composition of the samples, we also tried to measure the time of spin-lattice relaxation $T_1$ in addition to $T_2$ and $D$. Unfortunately, $T_1$ appeared to be insensitive to the presence of liquid-like inclusions in the hcp matrix.

Thus, the investigation of fast-grown solid helium (1\% $^3$He-$^4$He) samples by the NMR technique and measurement of the diffusion coefficient and the spin-spin relaxation time enabled us to identify growth defects formed in the crystals. The defects were mainly liquid droplets captured in the process of crystallization. The size of the droplets had a qualitative dependence proportional to the rate of cooling liquid helium  during its crystallization. The droplets disappear after annealing these two-phase samples. It is also shown that no liquid droplets were found in the samples grown at the cooling rate $\AC 0.08$~mK/s.

\subsection*{Acknowledgments}
This work has been partially supported by STCU Grant (Project \#5211).

\section*{References}

\bibliographystyle{iopart-num}
\bibliography{bib_disser_vekhov}

\end{document}